\documentclass[sigconf,10pt]{acmart}

\renewcommand\footnotetextcopyrightpermission[1]{} % removes footnote with conference information in first column
\usepackage{mathrsfs}
\usepackage{multirow}
\usepackage{bigstrut}
\usepackage{graphicx}
\usepackage{xspace}
\usepackage{fancyvrb}
\usepackage{subfigure}
\usepackage[inline]{enumitem}
\usepackage{xcolor}
\usepackage{caption}
\usepackage{times}
\usepackage{url}

\def\mprdma{Virtuoso}

\begin{document}

\title{A Novel Software-based Multi-path RDMA Solution for Data Center Networks}

\author{Feng Tian}
\affiliation{%
  \institution{University of Minnesota}
  \city{Minneapolis}
  \state{Minnesota}
}
\author{Wendi Feng}
\affiliation{%
  \institution{University of Minnesota}
  \city{Minneapolis}
  \state{Minnesota}
}
\author{Yang Zhang}
\affiliation{%
  \institution{University of Minnesota}
  \city{Minneapolis}
  \state{Minnesota}
}

\author{Zhi-Li Zhang}
\affiliation{%
  \institution{University of Minnesota}
  \city{Minneapolis}
  \state{Minnesota}
}

\begin{abstract}
	In this paper we propose {\em \mprdma}, a purely {\em software-based}  multi-path RDMA solution for data center networks (DCNs) to effectively utilize the rich multi-path topology for load balancing and reliability. As a ``middleware'' library operating at the user space, \mprdma~employs three innovative mechanisms to achieve its goal.  In contrast to existing hardware based MP-RDMA solution, \mprdma~can be readily deployed in DCNs with existing RDMA NICs. It also decouples path selection and load balancing mechanisms from hardware features, allowing DCN operators and applications to make flexible decisions by employing best mechanisms (as ``plug-in'' software library modules) as needed. Our experiments show that \mprdma~is capable of fully utilizing multiple paths with negligible CPU overheads. 
\end{abstract}


\maketitle

\section{Introduction}
Remote Direct Memory Access (RDMA) introduces the capability of  directly accessing the memory of a remote server by implementing the transport logic in hardware network interface cards  and bypassing CPU and kernel network stacks, thereby offering high bandwidth and low latency. Nowadays, RDMA is widely deployed over ``Converged'' Ethernet via RoCEv2 in modern data centers~\cite{rdmaml1,rdmaml2,rdmaml3}  to support machine learning and other data intensive applications. By design, RDMA is a point-to-point transport, where each RDMA connection is mapped onto a single network path. More specifically, RDMA operations ({\em verbs}) of an RDMA connection are transported along the same network path via single Queue Pair (QP); each message of an RDMA verb such as SEND, RECV, READ, WRITE is divided in segments of equal size and encapsulated in UDP packets, where the source and destination IP addresses of the UDP packets are set to those of the two communicating servers,  the destination port fixed at 4791 and the source port arbitrarily chosen. These are all done automatically by the RDMA NICs (or RNICs in short), which makes port number based path control mechanism as in MPTCP~\cite{mptcp} difficult in user space.

Data center networks (DCNs) are typically built using a ``spine-leaf'' topological structure with rich multiple paths, especially between spine routers, for load balancing and reliability~\cite{dcnarch,VL2}. As a point-to-point transport, RDMA does not take advantage of multiple paths in the underlying networks for load balancing and reliability~\cite{rdmautil1,rdmautil2,rdmautil3}. For machine learning and other data intensive applications, an RDMA read/write operation may involve remote transfer of a big chunk of data (``elephant flows''), which may not only take some time to deliver along a single path, but also cause congestion that can potentially affect ``mice flows'' from other applications, especially interactive applications with stringent latency requirements. MP-RDMA~\cite{mprdma} is the first work that attempts to address this limitation of existing RDMA (or rather, RoCEv2). It focuses on the challenges in implementing a multi-path RDMA solution in {\em hardware}, in particular, the limited memory resource in RNICs. By using source port to encode ``virtual path'' id (VP id) and influencing the path traversed by the RDMA UDP packets, it assumes and heavily relies on the underlying routers' ECMP mechanisms for load balancing among multiple paths. The proposed solution is emulated/prototyped using FPGA. As MP-RDMA requires replacing existing RNICs to new MP-RDMA capable NICs, it cannot be readily deployed in DCNs.

In this work, we propose and develop a purely {\em software-based multi-path} RDMA solution, dubbed {\em \mprdma{}}. Our solution employs {\em three key innovations}. 
First, we create multiple virtual interfaces -- each with a different (virtual) IP address of our choice -- and bind them to the same physical RNIC (effectively creating multiple virtual RNICs). Hence unlike MP-RDMA which manipulates the source port only, we control and manipulate source IP addresses of the RDMA UDP packets for load balancing and reliability. 
Second, we develop a user-space {\em middleware} layer which intercepts and split (large) messages of RDMA operations into multiple (smaller) messages, and dynamically maps them onto different paths at the sender side, and judiciously merge them together at the receiver side by passing them to the applications. Performing these operations correctly and incurring as little overheads as possible (especially, maintaining zero-copy) is nontrivial; it involves careful design and some clever tricks (see Section~\ref{sec:design}).
Third, we also implemented a user-space load balancer that consists of a congestion avoidance (for lossy network) and path probing component mechanism, to do a application aware load balancing.

\mprdma{}~offers several advantages over existing hardware-based multi-path RDMA solutions. As a purely software-based solution, it can be readily deployed in DCNs at scale with existing RDMA NICs and works regardless of the number of physical RNICs installed on servers. In contrast to MP-RDMA which implements ``built-in'' path selection, congestion control and traffic distribution mechanisms in hardware and hinges on ECMP to perform multi-path routing, \mprdma{}~decouples these mechanisms from hardware features, and allows DCN operators/applications to make flexible decisions by employing best mechanisms (as ``plug-in'' software library modules) as needed. For example, one can explicitly manage multi-path routing by setting appropriate forwarding rules (based on source and destination IP addresses), e.g., through an SDN controller. \mprdma{}~allows them to guide traffic distribution decisions.  
Our experiments show that \mprdma~can fully utilize multiple paths with negligible CPU overheads.

\section{Motivation \& Related Work}
\label{sec:motivation}
\subsection{RDMA/RoCE Basics}
RDMA allows applications to directly access remote memory with zero-copying and low CPU involvement by implementing the transport logic in hardware RNICs. RDMA over Converged Ethernet v2 (RoCEv2) has been widely deployed in data center networks to support compute- \& data-intensive applications such as machine learning, as it provides low latency and high bandwidth with little CPU overheads. Normally, RDMA requires a lossless network where Priority-based Flow Control (PFC) and Explicit Congestion Notification (ECN) are usually configured to prevent packet losses by pausing traffic transport and throttle traffic at the source.

 RDMA is a message based, point-to-point transport, where RDMA messages are divided into segments and encapsulated in UDP packets that are transported along a single path. Applications connect with each other using send and receive  {\em Queue Pairs (QP)}. An application initiates RDMA operations (or {\em verbs}) by posting {\em Work Requests (WRs)} (or {\em Work Queue Element} (WQE)), e.g., SEND/RECV or WRITE/READ to the QP, which commands the RNIC to transfer data to the memory of a remote host. For each application, there is also one (or more) completion queue (CQ); upon completing a WR, a completion queue element (CQE) is delivered to CQ.

\subsection{Multi-paths in Data Centers \& RDMA}
The ``leaf-spine'' topology in modern Data Center Networks (DCNs) offers rich path diversity~\cite{spineleaf1, VL2, dcnarch}. Switches and routers employ {\em built-in} Equal-Cost Multi-path (ECMP) for routing  based on hashes of 5-tuple packet/flow headers ($\langle$\texttt{}\textit{src IP}, \textit{dst IP}, \textit{src port}, \textit{dst port}, \textit{protocol number}$\rangle$).  ECMP suffers several issues in practice~\cite{ecmp_issue,dcnarch}, e.g., it is less effective when the number of paths is large, and it cannot perform intra-flow load balancing for large elephant flows. Other (software-based) solutions such as Valiant Load Balancing  and ``customized'' multi-path routing algorithms (e.g., by setting up explicit flow rules~\cite{VL2,dcnarch,Portland}) provide DCN operators and applications more control over multi-path routing and load balancing.  We remark that congestion often occurs at the core layer of a DCN~\cite{conga}; large ``elephant'' flows generated by data-intensive machine learning applications further contribute to this problem. They not only prolong their own flow completion times (FCTs), but also adversely affect other applications. It is therefore desirable to split such ``elephant'' flows to enable ``intra-flow'' load balancing across multiple (core) paths~\cite{conga,flowcell}.

MP-RDMA~\cite{mprdma} is the first to address the challenge that RDMA/RoCE v2 cannot effectively take advantage of rich multiple paths in DCNs~\cite{rdmautil1,rdmautil2,rdmautil3}. It proposes a hardware-based solution with ``built-in'' path selection and congestion avoidance mechanisms. The key challenge it focuses on is the limited memory in  RNICs  (see also FaRM~\cite{179767}, LITE~\cite{Tsai:2017:LKR:3132747.3132762} and INFINISWAP~\cite{201565} that tackle similar hardware constraints). As a hardware-based solution, it cannot be readily deployed without upgrading RNIC. It also heavily relies on ECMP for multi-path routing and load balancing. 

%\begin{figure}[h!]
%	\begin{center}
%		\includegraphics[width=0.8\columnwidth]{fig/mpu.eps}
%		\caption{Impact of Elephant Flows and Intra-Flow Load Balancing} % caption for whole figure
%		\label{fig:motivation}
		
%	\end{center}
%\end{figure}

We therefore seek a purely software-based multi-path RDMA solution operating in the user space  that works with existing RNICs while maintaining zero-copying and incurring as little CPU overheads as possible. A key enabling idea of our proposed solution is to create multiple virtual NICs (vNICs) and bind them to the same hardware RNIC, thereby allowing multiple IP addresses to be assigned to the same RNIC. Our solution allows a single RDMA application to create multiple {\em virtual} RDMA connections that are mapped to different paths. This is different from existing efforts in {\em virtualizing} RNICs~\cite{rdmavir1,rdmavir2,freeflow,hyv} with the goal to allow multiple VMs/containers to share the same RNIC with some level of isolation.  Compared with ``built-in'' multi-path routing and load balancing mechanisms, we also believe that it is imperative to provide DCN operators and applications with  flexibility in multi-path routing and load balancing decisions. For example, it has been shown that global congestion avoidance and traffic scheduling~\cite{cc1,cc2,perry2015fastpass} are essential in DCNs, and applications are best aware of traffic load distribution for adaptive load balancing~\cite{applb}.  %cc2
Similarly, Avatar~\cite{Qiu:2018:TEF:3232565.3232636} aims at making RDMA transport on a single NIC to be efficiently shared by eliminating lock contention and providing fair data scheduling via WR multiplexing.

\section{\mprdma{} Design}\label{sec:design}
\subsection{Overview}
\begin{figure}[htbp]
	\centering
	
	\subfigure[Virtuoso Overview]{
		\begin{minipage}[t]{0.5\linewidth}
			\centering
			\includegraphics[height=4.2cm]{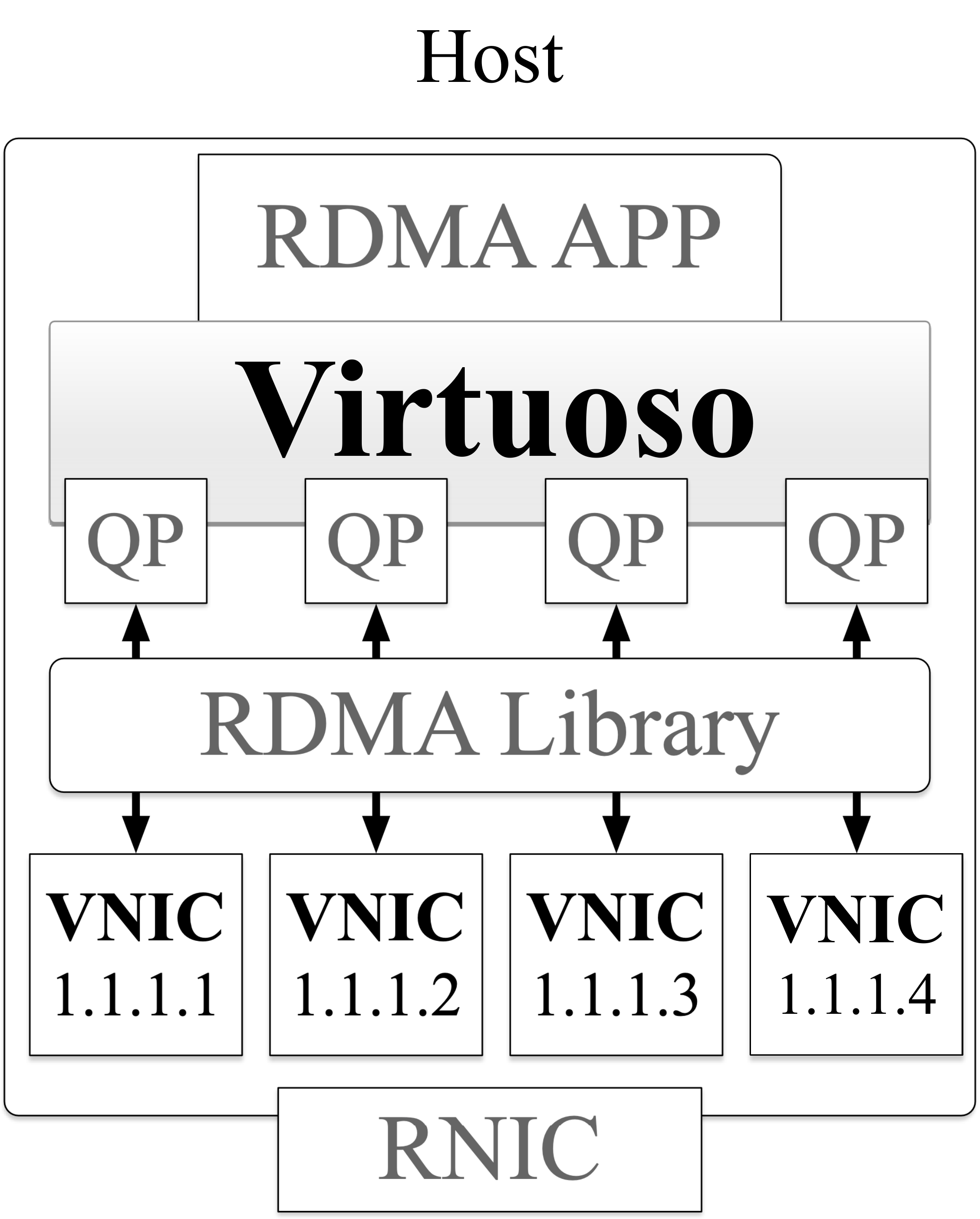}
			\label{fig:overview}
		\end{minipage}%
	}%
	\subfigure[System Design]{
		\begin{minipage}[t]{0.5\linewidth}
			\centering
			\includegraphics[height=4.2cm]{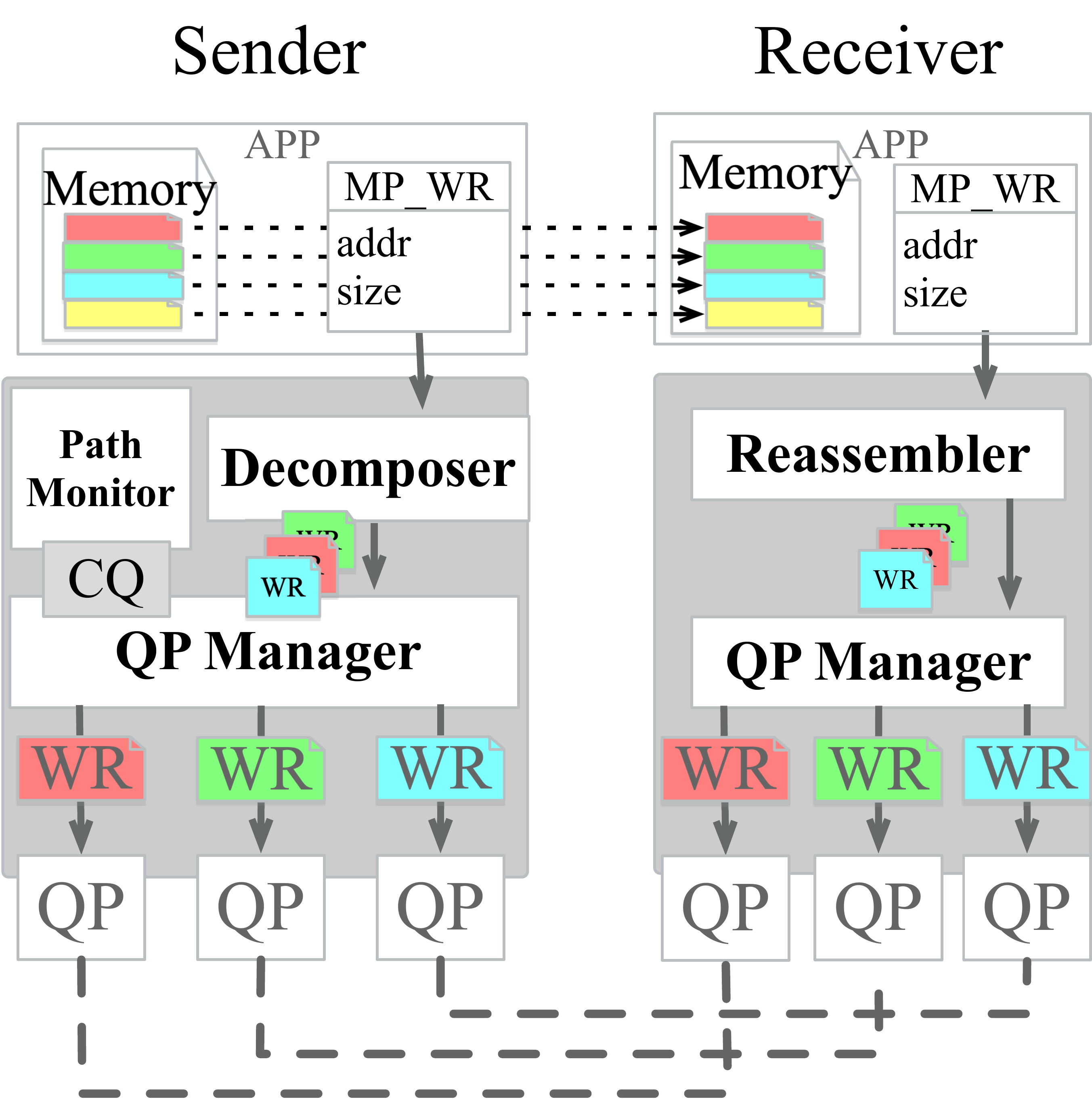}
			\label{fig:sys}
		\end{minipage}%
	}%
	\caption{\mprdma{}: Software Multi-path RDMA Solution  } % caption for whole figure
	\vspace{-0.3cm}
	\label{fig:design}
\end{figure}

%\label{sec:overview}
\mprdma{} is a software-based, modular multi-path RDMA framework. \mprdma{} sets up multiple virtual NICs (vNICs) on each physical RNIC using \texttt{IP alias}, each assigned with a distinct IP address (see Fig.~\ref{fig:overview}). In practice, RDMA uses a Global ID (GID) to identify each host, and RoCEv2 binds GIDs to the IP addresses of the interfaces using the IP table. Using vNICs, \mprdma{} is able to create multiple QPs using the standard RDMP libraries  \texttt{rdma\_cm} and \texttt{ibv\_verb}.  

%\begin{figure}[htbp]
%	\centering
	
%	\centering
%	\includegraphics[height=1.7in]{fig/design.eps}
%	\caption{Virtuoso Overview \& System Design} % caption for whole figure
%	\label{fig:design}
%	\vspace*{-0.3cm}
%\end{figure}

\mprdma{} maps each QP to a distinct {\em virtual} path (VP), and using the IP address associated with each vNIC as a {\em VP id}. As a middleware operating at the user space, \mprdma{} provides the same APIs (and RDMA verbs) as in the standard RDMA libraries, but prefixes them with the keyword \texttt{MP\_} as shown in Table \ref{fig:api}.  For example, an application invokes \texttt{MP\_connect()} to set up a \mprdma{} {\em multi-path} (logical) connection,  and uses \texttt{MP\_READ/SEND} and \texttt{MP\_WRITE/RECV} to post \mprdma{}   work requests (WRs), \texttt{MP\_WR}'s. On the sender side, \mprdma{} decomposes a large RDMA message (thereafter simply a ``flow'') contained in an \texttt{MP\_WR} into smaller ``sub-flows'', and distribute them to different QP's by generating the corresponding constituent WRs using the standard RDMA verbs. The sub-flows are ``merged'' at the receiver side. These are  illustrated in the right portion of Fig.~\ref{fig:sys}. \mprdma{} consists four major components, {\em QP Manager}, {\em Decomposer} (on the sender side), {\em Reassembler} (on the receiver side), and {\em Path Monitor \& Load Balancer}.

\begin{table}[hbt]
  \begin{tabular}{c|c}
    Standard RDMA API \& Verbs & \mprdma{} Version\\
    \hline
    rdma\_connect() & MP\_rdma\_connect()\\
    rdma\_disconnect() & MP\_rdma\_disconnect() \\
    ibv\_post\_send() & MP\_ibv\_post\_send() \\
    \hline
    WRITE/READ & MP\_WRITE/READ \\
    SEND/RECV & MP\_SEND/RECV\\
    
  \end{tabular}
  \caption{Interface \& Verb Design}
  \label{fig:api}
  \vspace*{-0.8cm}
\end{table}

\mprdma{} assumes that there is only one single port connection between ToR switch (but can also work with multiple ports) and RNIC while have multiple paths in the core layer of data center networks. The load balancing mechanism can be either ECMP (with known hash function) or static routing.

\subsection{\mprdma{} QP Manager}\label{sec:QP-manager}
\subsubsection{Transmission via Multiple QPs}
As discussed above, an RDMA application creates a (logical) multi-path connection using \mprdma{} APIs. \mprdma{} maps this logical connection to  multiple (virtual) paths by automatically setting up the corresponding QPs, one per path. To set up these QPs to work with the same application, we take advantage of several key features of RDMA. Recall that  in RDMA, memory must be registered before any RDMA verb can be post. The sender and receiver communicate and negotiate the address locations of the respective memory.  Each RDMA transport context (registered memory, QP) is maintained inside a Protect Domain (PD).  Inside this PD, these contexts can be shared and accessed by multiple QPs who within the same PD.

In order to associate the multiple QPs created by \mprdma{}  with the same application, the {\em QP manager} create them within the same PD. Furthermore, the same target memory region as specified by an RDMA application is also registered to this PD. This way the message in an \texttt{MP\_WRITE} or \texttt{MP\_SEND} can be transported through any of the QPs; in particular, for a large message, it can be divided into smaller chunks and transported via multiple QPs for load balancing. 

The advantage of this design is efficiency and flexibility: the QPs can concurrently manage the same memory region without memory copying and state transfer between PDs. This, however, creates a challenge at the receiver side when the two-sided \texttt{MP\_SEND} and \texttt{MP\_RECV} verbs are used: the receiver will not know in advance which QP the data will be arriving, thus which QP to post the corresponding \texttt{RECV} WR. We will discuss how this challenge is addressed in {\em Reassembler} of \mprdma{}, as well as how out-of-order (OOO) data is  handled in Section~\ref{sec:composer}. QP manager also creates a {\em shared} Completion Queue (CQ) for these QPs, so that it can poll this queue to query the CQEs of the WRs posted to any of these QPs. Note that each CQE has the corresponding WR information (e.g., WR id). Hence for each \texttt{MP\_WR} (a ``flow'') submitted by an application,  \mprdma{} can determine whether its constituent WRs (``sub-flows'') have been completed, thereby notifying the {\em Decomposer} to post a \texttt{MP\_WR} for application about the completion of the transmission task.

\subsubsection{MP\_connection \& MP\_disconnect}
In terms of connections between queue pair, it requires a transmission parameter (e.g., queue pair type (qp\_type) \& queue pair capabilities (max\_send\_wr)) exchange process, which involves several functions provided by standard RDMA libraries. This procedure works as the three-way hand shake procedure in TCP/IP. However, this procedure is handled in user space by application instead of the driver in kernel. Thus, \mprdma{} should handle all the parameter exchange tasks for multiple QPs. To simplify the connection procedure, \mprdma{} provides an uniformed interface, `\texttt{MP\_rdma\_connect()}', for multi-path connection which takes over the whole connecting procedure from application. Moreover, application can also configure these parameters by submitting configurations to \mprdma{}. The disconnection procedure of QPs is also similar and requires extra negotiation between two remote sides. Thus, \mprdma{} also provides an uniformed interface, `\texttt{MP\_rdma\_disconnect()}', for applications.

\subsection{\mprdma{} Decomposer}
The {\em Decomposer} component is responsible for WR generating, memory mapping and \texttt{MP\_CQE} generating. As the same as RDMA verbs, each \texttt{MP\_WR} (multi-path work request) contains the relevant metadata (memory location, size) regarding the target memory blocks it wants to access. At the sender side, the main task of {\em Decomposer} is to divide a large message  (``flow'') contained in a \texttt{MP\_WRITE} or \texttt{MP\_SEND} multi-path work request into smaller data chunks (``sub-flows''), and generate  the corresponding WRs (\texttt{WRITE} or \texttt{SEND}) for each sub-flow using the standard verb (\texttt{WRITE} or \texttt{SEND}).  Likewise, a \texttt{MP\_READ} WR that wants to access a large remote memory region (``flows'') will be divided at the sender side into multiple \texttt{READ} WRs, each accessing a smaller part of the target memory region (``sub-flows''). To facilitate the memory location and size matching between the sender and receiver, \mprdma{} divides the whole (application) memory space into blocks (this parameter is configurable). 

To decide the size of each sub-flows, the {\em Decomposer} will query the {\em Path Monitor \& Load Balancer}. Based on the path status, bandwidth and congestion information, {\em Path Monitor \& Load Balancer} provides a decision about the memory message and WR mapping where load balancing and congestion avoidance are considered (in section \ref{sec:pm}). Then, the {\em Decomposer} will generate WRs that maps different blocks of the memory, and pass them to {\em QP Manager}. After these WRs are successfully posted and completed, the {\em Decomposer} will be notified. Then it generates a corresponding \texttt{MP\_CQE} for entire message to notify the application of the completion.

\subsection{\mprdma{} Reassembler}
\label{sec:composer}
We first remark that while \mprdma{} performs the additional tasks of dividing a large message (``flow'') contained in an \texttt{MP\_READ}, \texttt{MP\_WRITE} or \texttt{MP\_SEND} into smaller messages (``sub-flows'') by generating a sequence of WRs. These WRs are distributed across multiple QPs, and are performed using the standard RDMA verbs (\texttt{READ}, \texttt{WRITE} or \texttt{SEND}). In other words, the RNIC will directly read/write  the corresponding data from or into the remote memory area in application's memory region as indicated by the verbs. Hence, \mprdma{} incurs {\em no additional memory copying}. 

\textbf{Out-of-Order (OOO)} is a common issue in multi-path transport, due to parallel transmission and variant delay on multiple paths. \mprdma{} leverages the benefit of direct memory wiring to resolve the OOO issue by buffering correctly received data into application memory. Once the data traffic arrived on the remote side, we have to merge sub-flows to reconstruct original memory region for receiver application. Since sub-flow traffic pay-loads are written to the memory directly by NIC hardware. The data flow will be composed correctly directly in the user space memory once we post the correct WR into the receive queue of \texttt{MP\_SEND/RECV} case (to identify the target memory addresses for each sub-flow); into the send queue of \texttt{MP\_WIRTE/READ} case (where the receiver side is totally passive).  When \mprdma{} uses \texttt{WRITE/READ} verbs as instructed by application submitted \texttt{MP\_WR}, the receiver side is totally passive (which means receiver requires no action after memory registration). Once the access key of remote memory is acquired by sender side, \mprdma{} can treat remote target memory as its own memory space without receiver's reaction for any transmission.

\begin{figure}[h!]

	\begin{center}
		\includegraphics[width=\columnwidth]{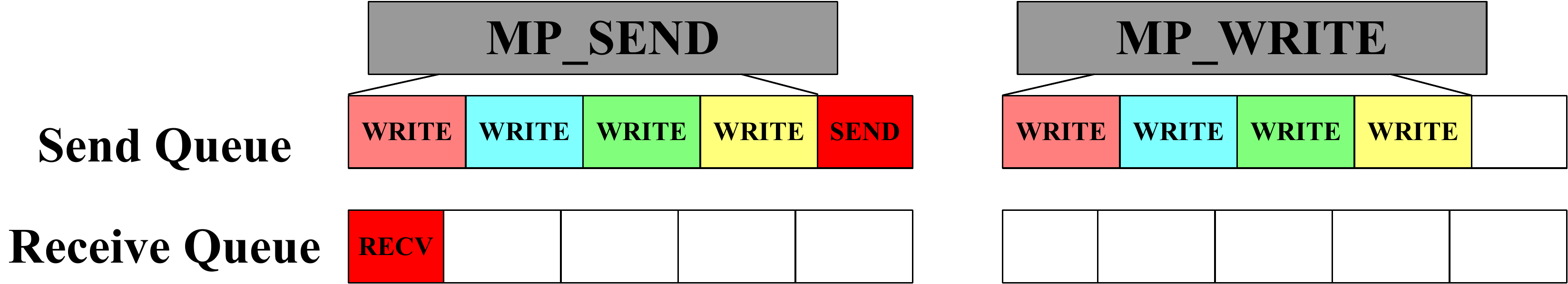}
		\caption{SEND/RECV \& Out-of-Order}
		\label{fig:sendwrite}
	\vspace*{-0.3cm}
	\end{center}
\end{figure}

 \textbf{MP\_SEND/RECV} verbs is a special case of OOO. Originally in RDMA,  each \texttt{SEND} consumes a \texttt{RECV} in receive queue. Moreover, \texttt{RECV} (who instructs RNIC to write data to the target memory address) is supposed to be posted before \texttt{SEND}'s arriving, which means the target addresses need to be determined in advance. However, the arriving order and data size of each sub-flow is unpredictable. So we cannot simply generate multiple \texttt{SEND/RECV} WRs as in \texttt{MP\_WRITE} case. So we propose a hybrid solution by combining \texttt{SEND/RECV} and \texttt{WRITE}. As illustrated in Fig~\ref{fig:sendwrite}, \texttt{WRITE} verbs who requires no \texttt{RECV}, are used to avoid beforehand memory address determination on receiver side. 
 
 Additionally, two-sided \texttt{SEND/RECV} needs to notify the application of accomplishment by posting a CQE into CQ (\texttt{MP\_CQE} in our case). However, one-sided \texttt{WRITE} verb cannot generate CQEs on receiver side. To this end, \mprdma{} posts an extra \texttt{RECV} to the receiving queue for receiver notification purpose. And also \mprdma{} appends an extra \texttt{SEND} after \texttt{WRITE} WRs to consume this \texttt{RECV}. Both the \texttt{RECV} and \texttt{SEND} are empty WRs (did not map any memory). As a result, when all the \texttt{WRITE} and \texttt{SEND/RECV} WRs are accomplished, CQEs will be posted to CQs on both sender and receiver sides. After polling the CQ, \mprdma{} can post a \texttt{MP\_CQE} to notify application using the metadata in CQE. 
 
For efficiency, we classify the \texttt{MP\_SEND} into two categories, small message and large message. For small message, a single \texttt{SEND} is used to send entire small message via arbitrary single path; for large message, the hybrid solution is used to load balance the elephant flow of the message onto multiple paths. 
 
\subsection{\mprdma{}~Path Monitor \& Load Balancer}
\label{sec:pm}
\begin{figure}[htbp]
	\centering
	\vspace*{-0.4cm}
	\subfigure[Lossless Network]{
		\begin{minipage}[t]{0.5\linewidth}
			\centering
			\includegraphics[height=2.2cm]{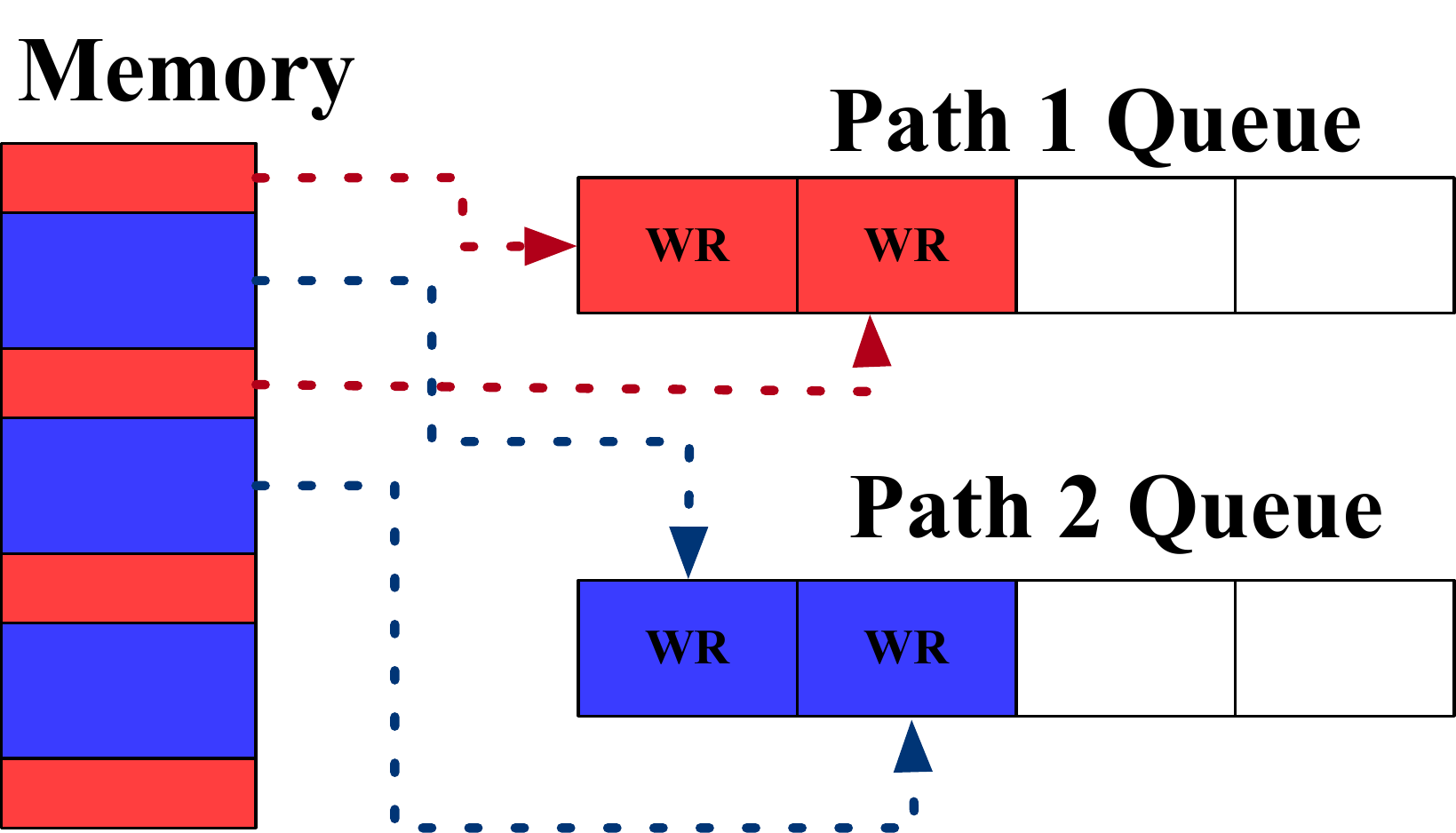}
			\label{fig:lossless}
		\end{minipage}%
	}%
	\subfigure[Lossy Network]{
		\begin{minipage}[t]{0.5\linewidth}
			\centering
			\includegraphics[height=2.2cm]{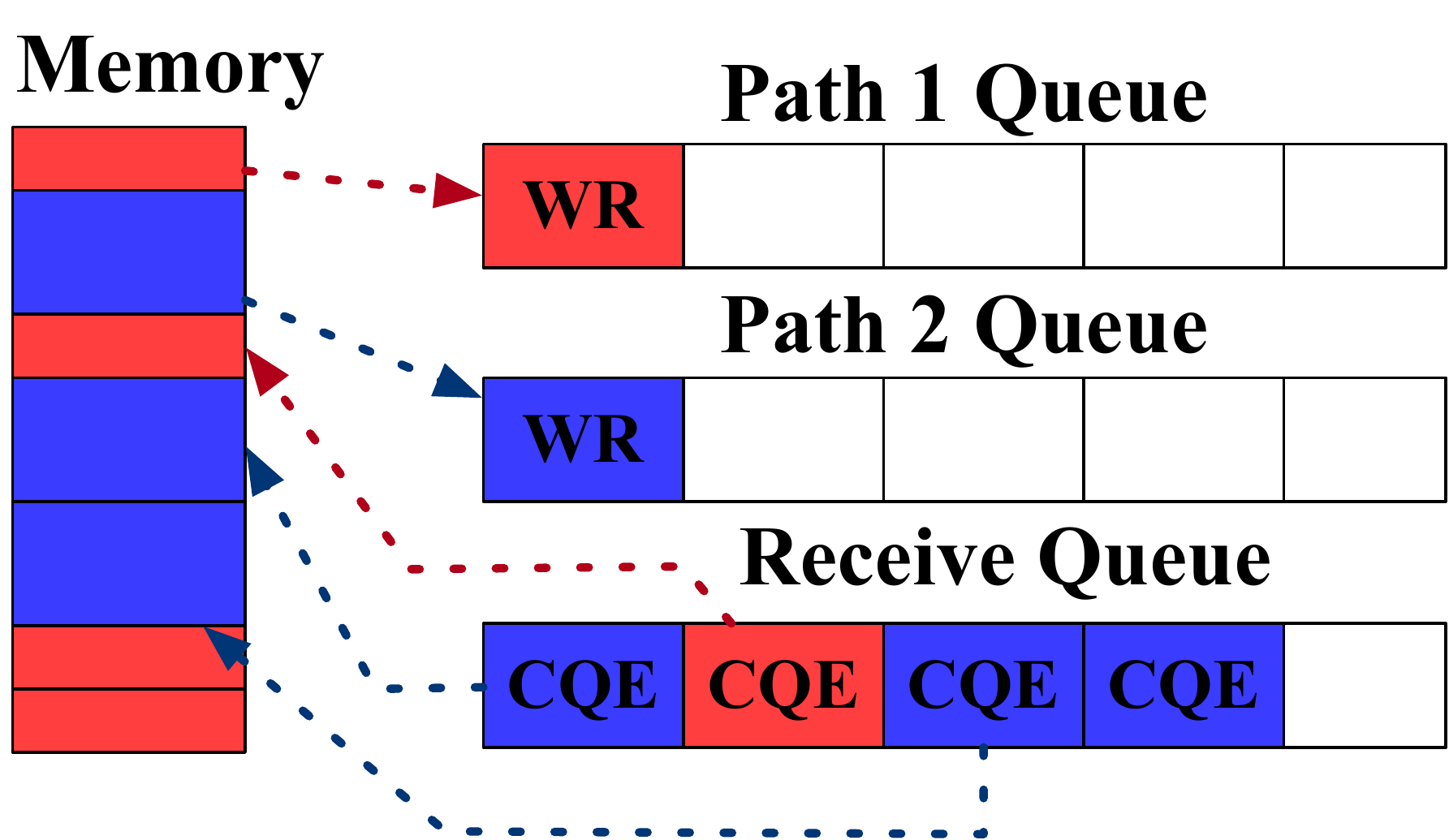}
			\label{fig:lossy}
		\end{minipage}%
	}%
    \vspace*{-0.3cm}
	\caption{User-space Load Balancing} % caption for whole figure
	\vspace*{-0.5cm}
	\label{fig:lb}
\end{figure}

\textbf{Load Balancing} is an essential task in multi-path transmission. \mprdma{} employs a pre-allocation mechanism to fit the RDMA verbs scenario. First, \mprdma{} probes the path capacity (e.g., bandwidth) using historical information (or other performance tools such as iPerf~\footnote{https://iperf.fr/iperf-download.php}). In current implementation, \mprdma{} initiates multiple probing flows (at least 512 KB) to estimate the capacity of each path by monitoring the flow completion times. Second, \mprdma{} distributes the incoming large data traffic into multiple sub-flows as follows.
\begin{equation}
\left\{
    \begin{array}{lr}
    \frac{data_{path_1}}{cap_{path_1}}=\frac{data_{path_1}}{cap_{path_1}}=\dots=\frac{data_{path_n}}{cap_{path_n}} \\
    \sum_{i=0}^{n}data_{path_i}=data_{total}
    \end{array}
\right.
\end{equation}
Here $cap_{path_i}$ and $data_{path_i}$ denote the estimated bandwidth and allocated data size for path $i$, respectively.. Then \mprdma{} maps the memory into WRs and submits them to QPs in Round-Robin scheduling as shown in Fig.~\ref{fig:lossless}. Current design is based on the assumption that the status of core paths are stable in short period. Since load balancing is totally decoupled from other components, more real-time and fine-grain load balancing mechanisms in user space will be explored in the future work.
 
\textbf{Congestion Avoidance} is also required in per sub-flow transmission. For instance, if the RNIC has insufficient resilient capability (e.g., Mellanox ConnectX-3 Pro) while the network is not well configured (lossy), mapping a large amount memory into a single WR (where RNIC transmits the data too fast) will cause packet loss in core switches (where the network bottleneck locates at). To resolve this, \mprdma{} limits the maximum trunk size of each WR using a congestion window based mechanism. Initially, \mprdma{} probes the threshold value of the trunk size of each sub-flow by binary increasing the chunk size while monitoring the shared CQ.  If a congestion happens (usually indicated by a CQE with \texttt{IBV\_WC\_RETRY\_EXC\_ERR} error code). \mprdma{} will decrease the chunk size to previous value and to find a maximum threshold in linear increasing. Moreover, WR construction and posting is also slightly different in lossy network. To avoid packet loss, \mprdma{} uses multiple WRs to map the sub-flow message of each path. The maximum chunk size value is used to determine the number of WRs. And also, these WRs will be posted in turns followed by success CQEs as shown in Fig.~\ref{fig:lossy}.
\section{Evaluation}
\label{sec:evaluation}
In this section, we introduce the implementation and evaluation of \mprdma{}. We evaluate the performance of \mprdma{}, and validate that \mprdma{} can fully utility multiple paths in the core of DCN with minimal CPU overhead.
\subsection{Implementation}
\mprdma{} is implemented as a user space ``middleware'' library on top of the standard RDMA libraries, \texttt{ib\_verbs} and \texttt{rdma\_cm}. \mprdma{} contains approximately 1500 lines of code (LoC) in C language.   \mprdma{} uses a thread-free method and event based mechanism  to handle multiple QPs establishment and data transmission.
An RDMA applications invokes \texttt{MP\_connect()} to create QP connections, and use \texttt{MP\_WRITE()/SEND()} to initiate a multipath data transmission. 
Additionally, we implemented two basic modules for congestion control and load balancing. However, they could be replaced easily for apps' own design.
\subsection{Testbed Setup}
\begin{figure}[h!]
	\begin{center}
	\vspace*{-0.2cm}
		\includegraphics[width=\columnwidth]{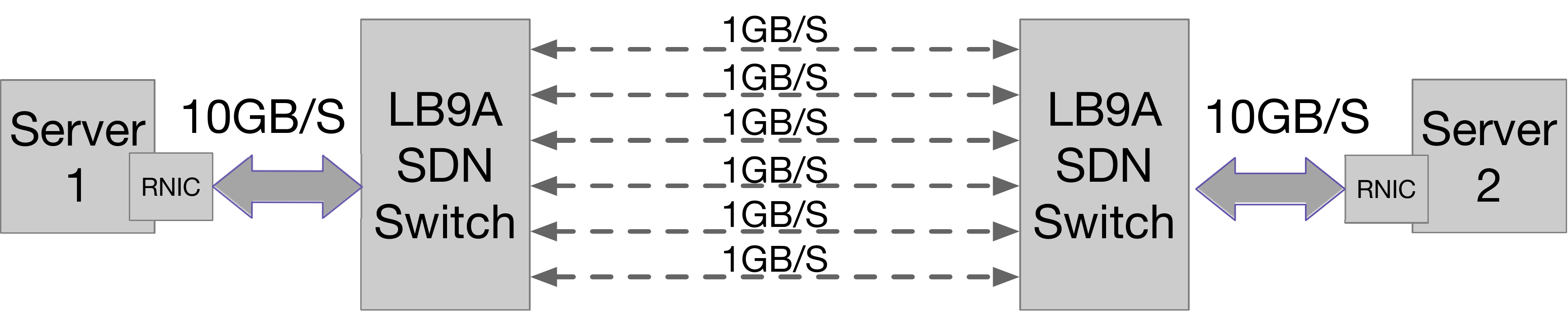}
		\caption{Testbed Setup} % caption for whole figure
		\label{fig:testbed}
	\end{center}
\end{figure}

Our testbed consists of two servers connected to two Top of Rack (ToR) switches with
multiple links between them to emulate the multipath scenario in spine-leaf DCN topology. 
The end-host server is Dell PowerEdge R430 with Intel Xeon CPU E5-2620v3@2.40GHz CPUs and 64GB RAM.  They are equipped with Mellanox ConnectX-3 40Gbps RNICs with the {\em MLNX\_OFED\_LINUX-4.6-1.0.1.1} driver with 10GB port enabled. The ToR switches are QuantaMesh\ T1048\ LB9A (SDN switch) to perform an ip based path mapping as shown in Fig. \ref{fig:testbed}. 
%Two paths are established between these two switches. 

\subsection{Multi-Path Utilization}
In this experiment, we evaluate the capability of \mprdma{} in path utilization. We can proof that \mprdma{} can fully utilize multiple paths to improve bandwidth in the network between ToR switches (core portion).

\begin{figure}[h!]
	\begin{center}
	\vspace*{-0.2cm}
		\includegraphics[width=\columnwidth]{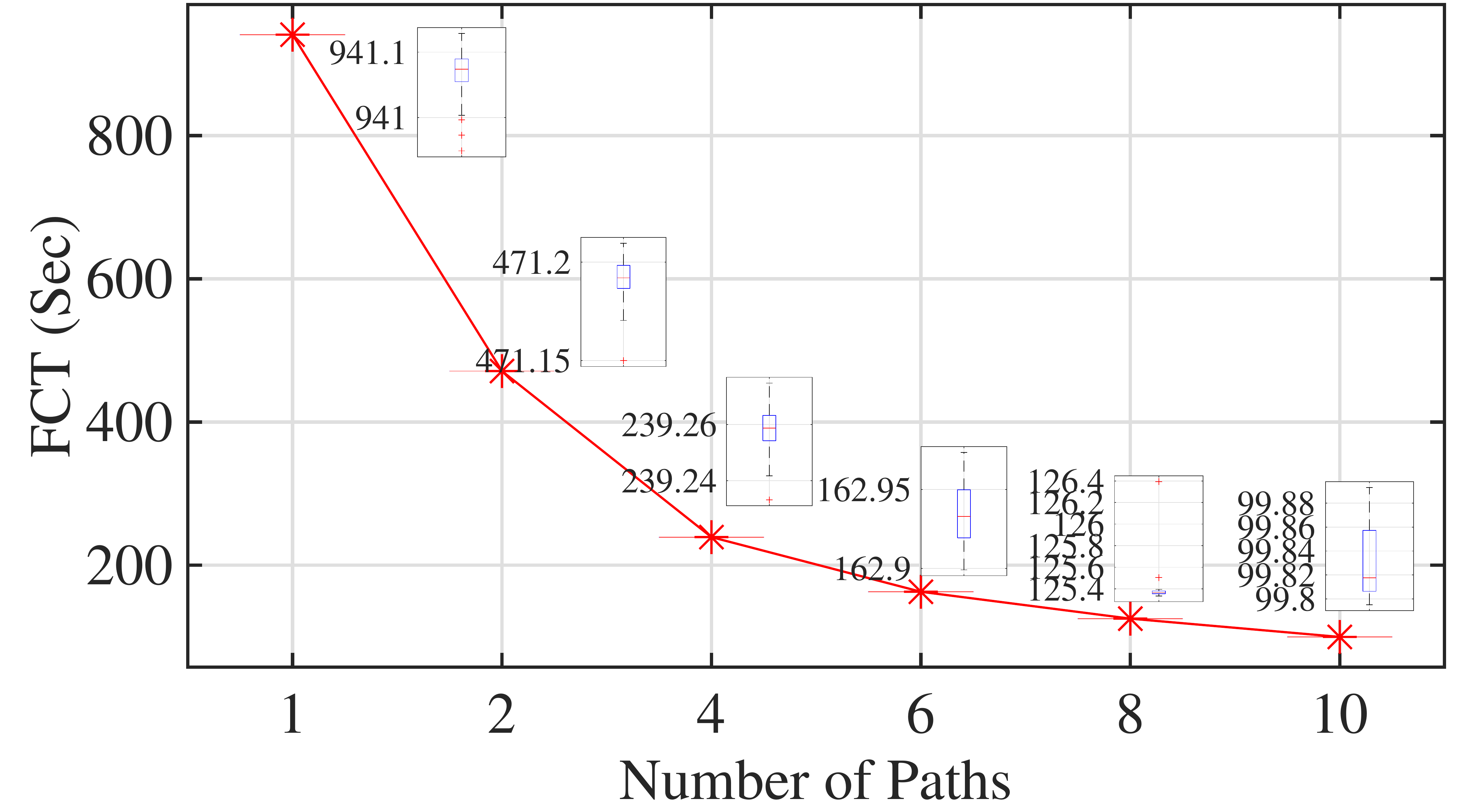}
		\vspace*{-0.3cm}
		\caption{Multiple Path Utilization (100GByte Flow)} % caption for whole figure
		\label{fig:100gb}
		\vspace*{-0.5cm}
	\end{center}
\end{figure}

Flow Completion Time is the matrix that we are using to evaluate the performance of \mprdma{} in using different number of paths(1, 2, 4, 6, 8 and 10). For each link between ToR switches, we limited the speed to 1GB/s while links between ToRs and servers are 10GB/s which introduces a bottleneck in core portion. As shown in Fig.~\ref{fig:100gb}, with the increasing of the number of used paths, the FCT will decrease obviously. And also, the benefit of using more paths can be leveraged under different sizes (from 10 MBytes to 100GBytes) of message size scenarios as shown in Fig.~\ref{fig:ms}, which means \mprdma{} can utilize multiple paths for better transport.

\begin{figure}[h!]
 \centering
 \vspace*{-0.3cm}
		\includegraphics[width=\columnwidth]{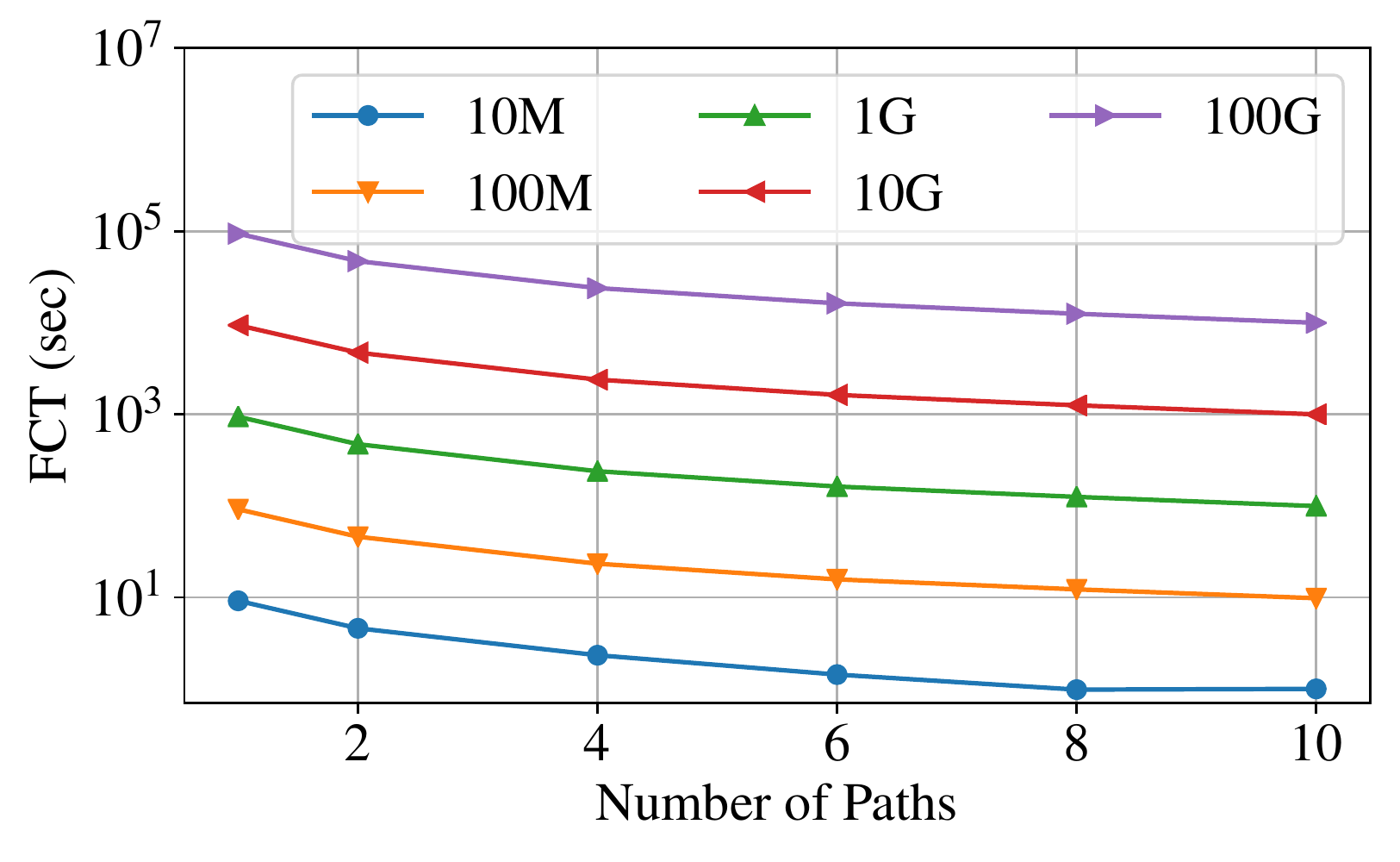}
		\caption{Different Flow Size Comparison} % caption for whole figure
		\label{fig:ms}
		\vspace*{-0.5cm}
\end{figure}

\subsection{Trunk Size \& Congestion Avoidance}
As in congestion control, if a WR submitted too much data at once, congestion will happen in bottleneck core network. Thus, utilizing multiple paths will potentially increase the trunk size that a WR can submit. As shown in Table.~\ref{fig:cs}, by using more paths, the trunk size can also increase. As a result, for a fixed size of data flow, we could save more CPU time by sending more data each single iteration.

Moreover, as shown in Table.~\ref{fig:cs}, using 2 or 4 paths can cause a decreasing of average trunk size compared with single path scenario. The reason is that the capabilities of limited number of extra paths still cannot patch the gap between core networks and RNIC. However, with more paths are used, the average trunk size of each can also be increase with less data allocated on each path in each iteration.

\begin{table}[hbt]
  \begin{tabular}{c|cc}
    \textit{\#} of Paths & Max Size (Byte) & Avg Size (Byte)\\
    \hline
    1& 700417 & 700417 \\
    2& 1300482 & 650241\\
    4& 2537172 & 634293\\
    6& 4405660 & 734280\\
    8& 9371656 & 1171457\\
    9& 18984993 & 2109400
  \end{tabular}
  \caption{Lossy Network Chunk Size Comparison}
  \label{fig:cs}
  \vspace*{-0.8cm}
\end{table}

\subsection{Multiple Flow \& Load Balancing}
As discussed in Section~\ref{sec:design}, multi-path transport can also increase the fairness by avoiding elephant flows blocking the mice ones. To validate that, we generate consistent data flow as background traffic while a mice flow (256 KBybe) is initiated in every 2 seconds. \mprdma{} splits the background elephant flow among 10 paths to avoid its blocking on single path. In comparative situation, \mprdma{} does not split the elephant flow, while mice flows are sharing the same path used by the elephant flow. Then, we compare the FCT of mice flows with/without load balancing of \mprdma{}.

As shown in Fig.~\ref{fig:ms}, with \mprdma{} splitting the elephant flow on multiple paths, the FCT of these mice flow will decrease due to extra bandwidth. In single path scenario, which is also the case without \mprdma{}'s load balancing, the background traffic occupies the shared single path and blocks the mice flows. As a result, the FCT of mice flows are increased.

\subsection{CPU Overhead}
We use CPU usage time (CPU cycles) to evaluate the CPU overhead of \mprdma{}. In this experiment, we tag the code in different points (e.g., the end of the \texttt{MP\_rdma\_connect()} function) to measure the CPU cycles used by different parts. The standard C library \texttt{time} is used to log the CPU clock of a specific time. 

Moreover, to avoid CQ polling caused extra CPU usage, we use event based completion queue polling (\texttt{ibv\_get\_cq\_event()} where the application will be blocked during data transmission). In this way, we could avoid the deviation caused by unnecessary CPU usage and measure only critical CPU overhead.

\begin{figure}[h!]
	\begin{center}
		\includegraphics[width=\columnwidth]{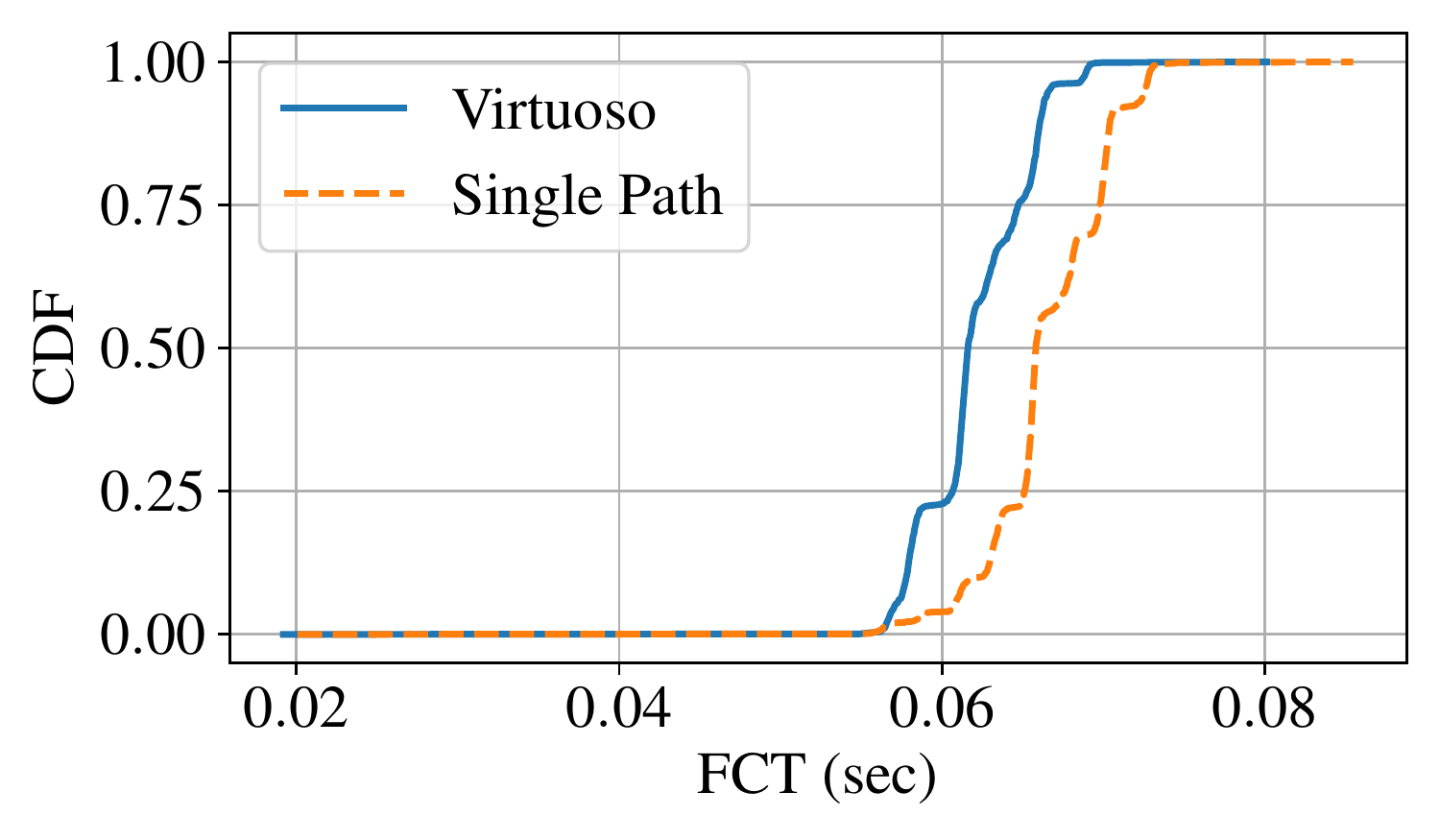}
		\vspace*{-0.5cm}
		\caption{Multiple Flows Interactions with \mprdma{}} 
		\label{fig:ms}
	\end{center}
	\vspace*{-0.3cm}
\end{figure}

\begin{figure}[h!]
	\centering

		\includegraphics[width=\columnwidth]{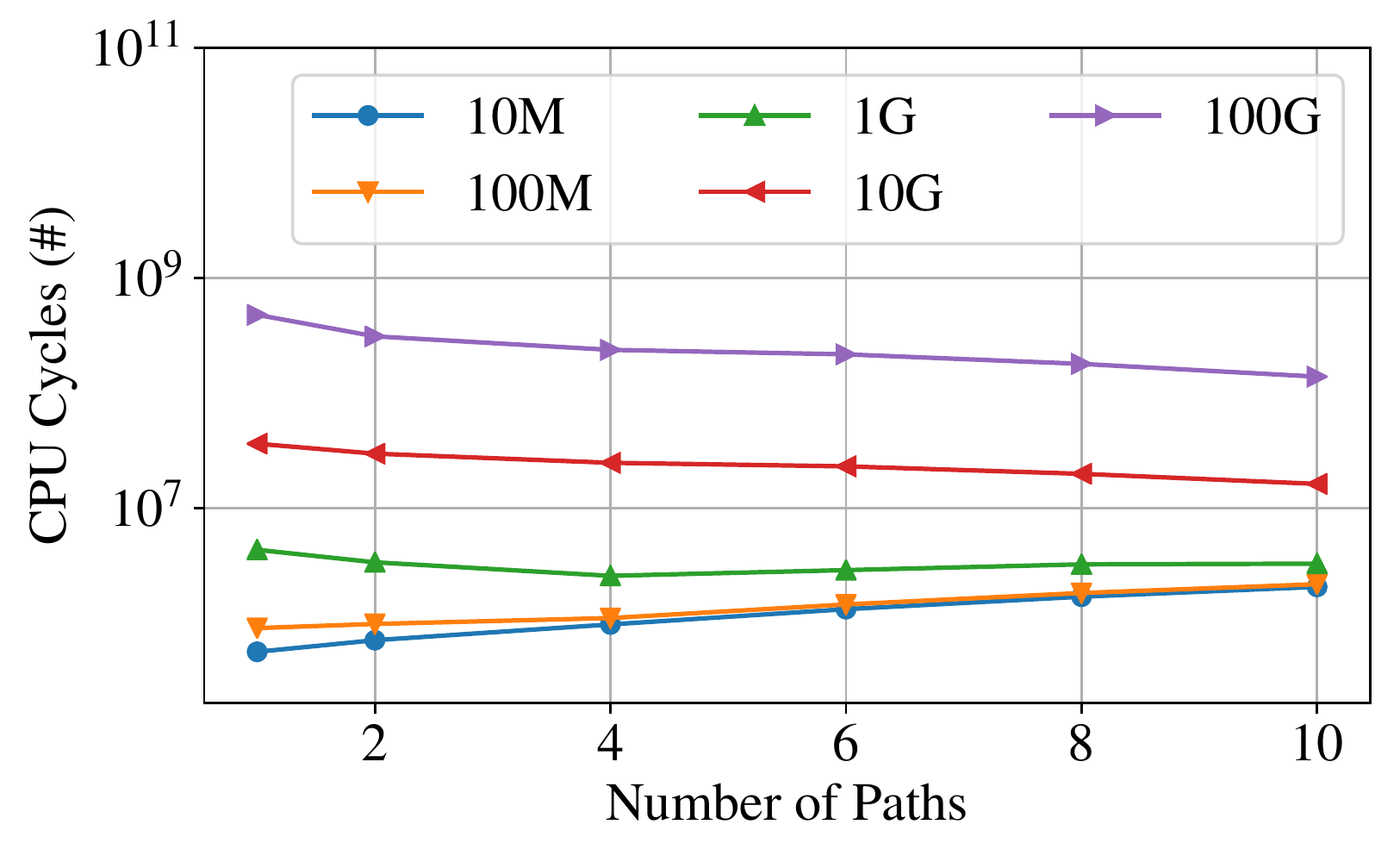}
		\vspace*{-0.5cm}
		\caption{CPU Usage Overhead Comparison} % caption for whole figure
		\label{fig:cpu_total}
		\vspace*{-0.3cm}
\end{figure}

As shown in Fig~\ref{fig:cpu_total}, with increasing the number of used paths especially in small message size scenarios, more CPU cycles are used in user space computation. However, large data size actually eliminates this side effects by increasing both bandwidth and trunk size to decrease the iterations for transmitting the same amount of data. Hence as a comprehensive conclusion, large data message should always leverage multiple paths while small data messages can use \mprdma{} to steer the flows to avoid congested paths.  

\section{Conclusion \& Future Work}

This paper presents {\em \mprdma}, a purely {\em software-based}  multi-path RDMA solution for data center networks which effectively utilizes multiple paths for load balancing and reliability. \mprdma{} employs VNICs to help RDMA applications split large flows into multiple smaller sub-flows and dispatch them among multiple paths to achieve user space load balancing. \mprdma{} can improve the bandwidth in core of DCNs by utilizing multiple paths but introduces negligible CPU overhead.

\mprdma{} is presented to bring inspirations for the community to leverage the flexibility of software visualization techniques. We plan to further 
\begin{enumerate*}[label=\roman*)]
\item provide a fine-grained yet efficient congestion control mechanism to achieve fast and dynamic load balance reaction and 
\item migrate real-world applications, like distributed TensorFlow~\cite{rdmaml2}, to evaluate the benefits of \mprdma{}, and to benefit the machine learning community.
\end{enumerate*}

%To address the challenge of coarse-grain load balance in RDMA deployed data center networks core layer using ECMP, we propose \mprdma{}, a novel virtual and dynamic multi-path RDMA solution to address challenges of user space path mapping and end-host level traffic splitting. Given design of \mprdma{}, applications could provide the core layer routing, e.g., ECMP, with fine-grain load balancing  capability. \mprdma{} employs the VNICs based method to help Applications spread the well know large traffic flow into multiple smaller identifiable sub-flow, to achieve fine-grain load balancing and congestion avoidance. Moreover, centralize transport scheduling mechanism could leverage our solution to achieve zero queue transport in RDMA data center due to the user space design of \mprdma{}.

%In future work, we will address the following challenge. 1) First, we have to provide a fine-grain congestion control mechanism to provide a faster load balancing reaction. 2) Second, we need to find a method for machine learning system such as distributed TensorFlow to achieve a better performance in the whole machine learning data center network.


\bibliographystyle{ACM-Reference-Format} 
\bibliography{main}

\end{document}